\documentclass[journal,letterpaper]{IEEEtran}

\usepackage{graphicx,cite,epsfig,amssymb,amsmath,amsfonts,multicol,subfigure,mathtools,bm,mathrsfs,setspace}

\begin{document}
\title{\huge Cocktail BPSK: Cross Power Utilization for High Data Rates}
\author{
	Bingli~Jiao,
	Yuli~Yang
	and~Mingxi~Yin
	\thanks{B. Jiao ({\em corresponding author}) and M. Yin are with the School of Electronics Engineering and Computer Science, Peking University, Beijing 100871, China (email: \{ jiaobl, yinmx\}@pku.edu.cn).}
	\thanks{Y. Yang is with the Department of Electronic and Electrical Engineering, University of Chester, Chester CH2 4NU, U.K. (e-mail: y.yang@chester.ac.uk).}
	
}

\maketitle

\begin{abstract}
This paper proposes a novel transmission strategy, referred to as cocktail BPSK, where two independent BPSK symbols are layered with various weights to be transmitted and, capitalizing on the cross power utilization, will be demodulated at the receiver without interference from each other. To evaluate the performance of the proposed scheme, its achievable data rate is formulated over additive white Gaussian noise (AWGN) channels. Based on the theoretical analysis, numerical results are provided for the performance comparisons between the proposed scheme and conventional transmission schemes, which substantiate the validity of the proposed scheme. Specifically, when the signal-to-noise power ratio (SNR) is small, the achievable data rate of the proposed scheme outperforms the channel capacity achieved by Gaussian-distributed inputs. 

\end{abstract}


\IEEEpeerreviewmaketitle

\section{Introduction}

Ideally, when the transmitted signals form a Gaussian ensemble, the AWGN channel capacity~\cite{Shannon1948}, i.e., the maximum data rate of the transmission over an AWGN channel, is achieved at
\begin{equation} \label{eqShannon}
C = \log_2 (1 + \rho)
\end{equation}
in the unit of [bits/sec/Hz], where $\rho$ is the received SNR.

In practice, the transmitted signals are finite-alphabet inputs, mapped by the information source through modulations, e.g., BPSK, QPSK, 8QAM, 16QAM, etc.  The data rates of finite-alphabet inputs can be expressed by logarithmic functions of their received SNRs as well~\cite{UngerIT, Blahut}.  Due to an important feature of the logarithm, when the SNR $\rho$ is small, logarithmic functions of $\rho$ approximate to linear functions of $\rho$~\cite{math}.  As such, the data rates achieved by finite-alphabet inputs approach the AWGN channel capacity, i.e., the case of the Gaussian-distributed input, in the low-SNR region.

Another important feature of the logarithm is its strict concavity, which leads to $\log_2(\rho) \leq \log_2(\rho_1) + \log_2(\rho_2)$ if $\rho = \rho_1 + \rho_2$.  Concerning this, in~\cite{jiao} we exploited multiple data streams to convey the information source for achieving higher data rate than that of a single data stream.  Since both Euclidean geometry and Hamming distance were involved in this conception, we have not sorted out a specific solution to address the complicated operations in previous works.

Tracking the conception of exploiting multiple data streams, we simplify the construction of this problem and focus on enabling parallel transmissions to achieve higher data rate in Euclidean geometry.  In the past decades, orthogonal codes have been exploited to carry out parallel transmissions in the Euclidean space~\cite{ortho1, ortho2}, where multiple data streams are combined in parallel transmissions over a single channel.  However, the sum rate of these data streams is not increased at all in comparison with the rate of a single data stream, because the orthogonality in Euclidean geometry isolates the resource(s) allocated among these data streams.

On the other hand, the relaxation of non-orthogonal combination provides more degrees of freedom for the system design to improve the transmission rate.  Even so, the simultaneous transmissions of multiple non-orthogonal data steams result in inter-stream interference and, accordingly, prevent the data rate gain.

Bearing the natures of both orthogonal and non-orthogonal combinations in mind, in this paper we embed the cross power utilization into weighted sum and propose a novel transmission strategy, referred to as cocktail BPSK, which allows the layering of two independent BPSK symbol streams with various weights to be demodulated without the inter-stream interference, thanks to the cross power utilization.

To detail the proposed scheme and its performance, the
remainder of this paper is organized as follows. Section II
prescribes the basic strategy of cocktail BPSK, i.e., one-dimensional cocktail BPSK, and its generalization into a two-dimensional case.
In Section III, the analysis framework
of achievable data rate is established for cocktail BPSK over AWGN channels.  Based on the theoretical analyses,
Section IV demonstrates numerical results to compare the proposed scheme
with conventional transmission schemes and substantiate the validity
of the proposed scheme. Finally, this paper is concluded
in Section V.

Throughout the paper, the following mathematical notations are used: Vectors are denoted by boldface lowercase letters.  The real part and the imaginary part of a complex number are denoted by $\textrm{real}(\cdot)$ and $\textrm{imag}(\cdot)$, respectively. Moreover, $p(\cdot)$ denotes the probability density function (PDF) of a random variable, and $\mathcal{E} \{ \cdot \}$ represents the expectation (mean) operator.

\section{Cocktail BPSK}

In this section, the transmission strategy referred to as cocktail BPSK is proposed to transmit the information source through the layering of two independent BPSK symbol streams, aiming to achieve high data rate.  Herein, the basic strategy of the proposed scheme is firstly presented as one-dimensional cocktail BPSK and, then, generalized into two-dimensional cocktail BPSK to double the achievable data rate.

\subsection{One-Dimensional Cocktail BPSK}
\label{secTx}

Consider two independent BPSK-modulated symbol streams, $\textbf{x} = [x_1, x_2, \cdots]$ and $\textbf{z} = [z_1, z_2, \cdots]$, conveyed over a memoryless AWGN channel, where the BPSK symbols $x_k, z_k \in \{+1, -1\}$, $k = 1, 2, \cdots$.

With the proposed cocktail BPSK, the transmitter layers $\textbf{x}$ and $\textbf{z}$ with various weights.  That is, the transmitted signal in the $k^\textrm{th}$ symbol period is a weighted sum of $x_k$ and $z_k$, $k = 1, 2, \cdots$.  For simplicity, we shall omit the symbol period index $k$.

Without loss of generality, the received signal in an arbitrary symbol period is expressed as
\begin{eqnarray}
\begin{array}{l}\label{eqT1}
y = \textbf{w} \begin{bmatrix}
           x \\ z
         \end{bmatrix} + n,
\end{array}
\end{eqnarray}
where $y$ is the received signal and $n$ is the received AWGN component, i.e., a random variable from a normally distributed ensemble of power $\sigma_N^2$, denoted by $n \sim \mathcal{N}(0,\sigma_N^2)$.  The $1 \times 2$ row vector $\textbf{w} = [w_1, w_2]$ consisting of the two weights in the layering is defined by
\begin{equation}
\textbf{w} = \left\{ \,
\begin{IEEEeqnarraybox}[][c]{s?s}
\IEEEstrut
$[\alpha, 0]$ & if $x = z$, \\
$[0, \beta/2]$ & if $x \neq z$,
\IEEEstrut
\end{IEEEeqnarraybox} 
\right.
\end{equation}
where $\alpha$ and $\beta$ are two positive real numbers ($\alpha > \beta >0$).

In fact, the key to the proposed cocktail BPSK lies in the design of the two weights.  Table I details the 4 equiprobable cases of the transmitted signal.

\newcommand{\tabincell}[2]{\begin{tabular}{@{}#1@{}}#2\end{tabular}}
\begin{table}[htb]
	\renewcommand{\arraystretch}{1.33}
	\centering
	\small
	\caption{Equiprobable Transmitted Signals in Layered BPSK}
	\label{Table1}
	\begin{tabular}{c | c c |c}
		\hline
                     & & & \\ [-1.2em]
		\tabincell{c}{Case} & \tabincell{c}{$x$} & \tabincell{c}{$z$} & \tabincell{c}{$\displaystyle{\renewcommand{\arraystretch}{1} \textbf{w}\begin{bmatrix} x \\ z \end{bmatrix}}$}  \\
                     & & & \\ [-1.2em]
            	\hline
		\tabincell{c}{1} & \tabincell{c}{$+1$} & \tabincell{c}{$+1$} & \tabincell{c}{$\alpha$}  \\
		\hline
		\tabincell{c}{2} & \tabincell{c}{$-1$} &\tabincell{c}{ $-1$} & \tabincell{c}{$-\alpha$}  \\
		\hline
		\tabincell{c}{3} & \tabincell{c}{$+1$} & \tabincell{c}{$-1$} & \tabincell{c}{$-\beta/2$} \\
		\hline
		\tabincell{c}{4} & \tabincell{c}{$-1$} & \tabincell{c}{$+1$} & \tabincell{c}{$\beta/2$}  \\
		\hline
\end{tabular}
\end{table}

Upon receiving $y$, the receiver will firstly demodulate the BPSK symbol $z$ in the same way as the demodulation of conventional BPSK: If $y > 0$, we choose $\hat{z} = +1$; otherwise, $\hat{z}= -1$.

Then, the BPSK symbols $x$ will be demodulated by a conventional BPSK demodulator as well, based on the demodulated symbol $\hat{z}$.  In detail, the signal sent to the BPSK demodulator is
\begin{equation} \label{eqRx}
\tilde{x} = y - \hat{z} \beta.
\end{equation}
Subsequently, we have $\hat{x} = +1$ if $\tilde{x}>0$ or $\hat{x}= -1$ if $\tilde{x}<0$.

\subsection{Two-Dimensional Cocktail BPSK}

Now, we generalize the basic cocktail BPSK into a two-dimensional scenario, by introducing another dimension composed of BPSK symbols $\{+j,-j\}$, where $j = \sqrt{-1}$ stands for the imaginary unit.  In this dimension, there are two independent BPSK-modulated symbol streams as well, denoted by $\textbf{x}' = [x'_1, x'_2, \cdots]$ and $\textbf{z}'=[z'_1, z'_2, \cdots]$, where $x'_k \in \{+j, -j\}$ and $z'_k \in \{+j, -j\}$, $k = 1, 2, \cdots$.

Moreover, $\textbf{x}'$ and $\textbf{z}'$ are layered in the same way as $\textbf{x}$ and $\textbf{z}$ are layered, with various weights.  As the dimension composed of $\{+j,-j\}$ and the dimension composed of $\{+1, -1\}$ are orthogonal, the cocktail made with $\textbf{x}'$ and $\textbf{z}'$ will be transmitted together with the layering of $\textbf{x}$ and $\textbf{z}$.

Without loss of generality, in an arbitrary symbol period, the received signal of the two-dimensional cocktail BPSK is expressed as
\begin{eqnarray}
\begin{array}{l}\label{T2m1}
y'=\textbf{w} \begin{bmatrix}
           x \\
           z
         \end{bmatrix} + \textbf{w}' \begin{bmatrix}
           x' \\
           z'
         \end{bmatrix} + n',
\end{array}
\end{eqnarray}
where $y'$ and $n' \sim \mathcal{CN}(0,\sigma_N^2)$ are the complex-valued signal and the complex AWGN component, respectively, observed at the receiver of the two-dimensional cocktail BPSK.

The $1 \times 2$ row vector $\textbf{w}'$, consisting of the two weights in the layering of $\textbf{x}'$ and $\textbf{z}'$, is designed as
\begin{equation}
\textbf{w}' = \left\{ \,
\begin{IEEEeqnarraybox}[][c]{s?s}
\IEEEstrut
$[\alpha', 0]$ & if $x' = z'$, \\
$[0, \beta'/2]$ & if $x' \neq z'$,
\IEEEstrut
\end{IEEEeqnarraybox} 
\right.
\end{equation}
where $\alpha'$ and $\beta'$ are two positive real numbers ($\alpha' > \beta' >0$).

At the receiver, the BPSK symbols $z$ and $z'$ will be firstly demodulated according to the real part and the imaginary part of the received signal $y'$, respectively.  Elaborating slightly further, we have 
\begin{equation}
\hat{z} = \left\{ \,
\begin{IEEEeqnarraybox}[][c]{s?s}
\IEEEstrut
$+1$ & if $\textrm{real}(y')>0$, \\
$-1$ & if  $\textrm{real}(y')<0$;
\IEEEstrut
\end{IEEEeqnarraybox} 
\right.
\end{equation}

\begin{equation}
\hat{z}' = \left\{ \,
\begin{IEEEeqnarraybox}[][c]{s?s}
\IEEEstrut
$+1$ & if $\textrm{imag}(y')>0$, \\
$-1$ & if  $\textrm{imag}(y')<0$;
\IEEEstrut
\end{IEEEeqnarraybox} 
\right.
\end{equation}

Then, by generating the signal $\tilde{x}' = y' - (\hat{z} \beta + \hat{z}' \beta')$, the BPSK symbols $x$ and $x'$ will be demodulated as the following:
\begin{equation}
\hat{x} = \left\{ \,
\begin{IEEEeqnarraybox}[][c]{s?s}
\IEEEstrut
$+1$ & if $\textrm{real}(\tilde{x}')>0$, \\
$-1$ & if  $\textrm{real}(\tilde{x}')<0$;
\IEEEstrut
\end{IEEEeqnarraybox} 
\right.
\end{equation}

\begin{equation}
\hat{x}' = \left\{ \,
\begin{IEEEeqnarraybox}[][c]{s?s}
\IEEEstrut
$+1$ & if $\textrm{imag}(\tilde{x}')>0$, \\
$-1$ & if  $\textrm{imag}(\tilde{x}')<0$;
\IEEEstrut
\end{IEEEeqnarraybox} 
\right.
\end{equation}


\section{Achievable Data Rate}
\label{DataRate}

As the demodulation of cocktail BPSK is exactly the same as that of conventional BPSK,  we shall start the establishment of the analysis framework with the achievable data rate of conventional BPSK that is characterized by a function of the BPSK amplitude $A$ and the AWGN power $\sigma_N^2$, expressed as~\cite{Shannon1948}
\begin{equation}\label{rateBPSK}
\begin{split}
& \ \mathbb{R}_\textrm{BPSK}(A,{\sigma_N^2}) = {\rm{H}}(Y) - {\rm{H}}(N) \\
& =  - \int_{-\infty }^{ + \infty } {p(y){{\log }_2}p(y){\rm{d}}y}  - {\log _2}(\sqrt {2\pi e{\sigma_N^2}} )
\end{split}
\end{equation}
where ${\rm{H}}(Y) = \mathcal{E} \{- \log_2 p(y)\}$ is the entropy of the received signal with the PDF of the received signal given by
\begin{equation}
p(y) = \frac{1}{2}\frac{1}{\sqrt{2\pi \sigma_N^2}}\left(e^{-\frac{(y-A)^2}{2 \sigma_N^2}} + e^{-\frac{(y + A)^2}{2 \sigma_N^2}}\right).
\end{equation}
Moreover, ${\rm{H}}(N) = {\log _2} (\sqrt{2 \pi e \sigma_N^2})$ is the entropy of the AWGN.

The achievable data rate of one-dimensional cocktail BPSK, i.e., the basic strategy of cocktail BPSK, denoted by $R_1$, can be expressed as
\begin{equation} \label{eqR1}
\mathbb{R}_1 = \mathbb{R}_Z + \mathbb{R}_X,
\end{equation}
where $\mathbb{R}_Z$ and $\mathbb{R}_X$ are the achievable data rate of the BPSK symbol streams $\textbf{z}$ and $\textbf{x}$, respectively.

As the demodulation of $\textbf{z}$ is equivalent to the demodulation of conventional BPSK with two equiprobable amplitudes, $\alpha$ and $\beta/2$, the data rate achieved by the BPSK symbol steam $\textbf{z}$ is calculated using 
\begin{eqnarray}
\begin{array}{l}\label{rateZ} \displaystyle
\mathbb{R}_z= \frac{1}{2} \mathbb{R}_\textrm{BPSK}(\alpha, \sigma_N^2) + \frac{1}{2} \mathbb{R}_\textrm{BPSK}(\beta/2, \sigma_N^2),
\end{array}
\end{eqnarray} 
where the function $\mathbb{R}_\textrm{BPSK}(\cdot, \cdot) $ is given by (\ref{rateBPSK}).

Similarly, the demodulation of the symbol stream $\textbf{x}$, given in (\ref{eqRx}), is equivalent to the demodulation of conventional BPSK with two equiprobable amplitudes, $(\alpha-\beta)$ and $\beta/2$.  Therefore, the achievable data rate of the BPSK symbol stream $\textbf{x}$ is obtained by 
\begin{eqnarray}
\begin{array}{l}\label{rateX} \displaystyle
\mathbb{R}_x= \frac{1}{2} \mathbb{R}_\textrm{BPSK}(\alpha-\beta, \sigma_N^2) + \frac{1}{2} \mathbb{R}_\textrm{BPSK}(\beta/2, \sigma_N^2).
\end{array}
\end{eqnarray}

For the purpose of comparison, the AWGN channel capacity at the point of $\rho = 0$ is approximated by the Taylor's expansion as 
\begin{eqnarray}
\tilde{C} \doteq \rho \log_2 e,
\end{eqnarray}
where $\log_2 e$ is the derivative of the AWGN channel capacity at $\rho=0$~\cite{EbN0}.

As shown in the Appendix, around the point of zero SNR, both the achievable data rates of finite-alphabet inputs and their derivatives are approximately equal to those of the AWGN channel capacity.  As a result, the achievable data rate of conventional BPSK at zero SNR can be estimated by
\begin{eqnarray}
\begin{array}{l}\label{EquTaylor} \displaystyle
\tilde{\mathbb{R}}_\textrm{BPSK} \doteq \rho_\textrm{BPSK} \log _2 e, 
\end{array}
\end{eqnarray}
where $\rho_\textrm{BPSK}$ denotes the received SNR in conventional BPSK.  In the performance comparison between one-dimensional cocktail BPSK and the conventional BPSK, the average received SNR of the former is constrained by the received SNR of the latter.  Thus, from Table \ref{Table1}, we have
\begin{eqnarray}
\rho_\textrm{BPSK} = \frac{1}{\sigma_N^2}\left[\frac{1}{2} \alpha^2 + \frac{1}{2} \left(\frac{\beta}{2}\right)^2 \right].
\end{eqnarray}

Based on (\ref{rateZ}) and (\ref{rateX}), (\ref{EquTaylor}) will lead to the achievable data rate of the cocktail BPSK approximated at zero SNR as
\begin{eqnarray}
\begin{array}{l}\displaystyle
\tilde{\mathbb{R}}_1 \doteq (\rho_z + \rho_x)\log _2 e, 
\end{array}
\end{eqnarray}
where
\begin{eqnarray} \label{rhoz}
\rho_z = \rho_\textrm{BPSK}
\end{eqnarray}
and
\begin{eqnarray} \label{rhox}
\rho_x = \frac{1}{\sigma_N^2}\left[\frac{1}{2} (\alpha-\beta)^2 + \frac{1}{2} \left(\frac{\beta}{2}\right)^2\right]
\end{eqnarray}
are the equivalent SNRs of $\textbf{z}$ and $\textbf{x}$, respectively, in the cocktail BPSK demodulation.

Apparently, the difference between $\tilde{\mathbb{R}}_1$ and $\tilde{\mathbb{R}}_\textrm{BPSK}$ lies in $\rho_x$.  That is,
\begin{eqnarray} \label{eqDiff}
\tilde{\mathbb{R}}_1 - \tilde{\mathbb{R}}_\textrm{BPSK} \doteq \rho_x \log_2 e
\end{eqnarray}
at the point of zero SNR.

In the case of two-dimensional cocktail BPSK, the data rate achieved by the layering of $\textbf{x}'$ and $\textbf{z}'$ is denoted by $\mathbb{R}'_1$, which can be obtained in the same way as the formulation of $\mathbb{R}_1$.  As a consequence, the data rate of two-dimensional cocktail BPSK, denoted by $\mathbb{R}_2$, is achieved at
\begin{equation}\label{eqR2}
\mathbb{R}_2 = \mathbb{R}_1 + \mathbb{R}'_1,
\end{equation}
where $\mathbb{R}_1$ is given by (\ref{eqR1}) and $\mathbb{R}'_1$ is obtained by simply replacing $\alpha$ and $\beta$ in the calculation of $\mathbb{R}_1$ with $\alpha'$ and $\beta'$, respectively.  If $\alpha=\alpha'$ and $\beta = \beta'$, we shall have $\mathbb{R}_2 = 2 \mathbb{R}_1$.

\section{Numerical Results and Discussions}

Herein, we illuminate numerically the performance of the proposed cocktail BPSK in the metric of achievable data rate and compare it with conventional transmission schemes.

\begin{figure}[!t]
	\centering
	\includegraphics[width=0.47\textwidth]{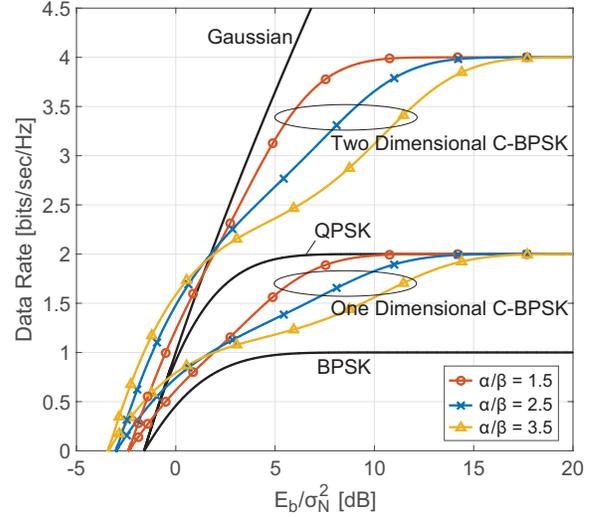}
	\caption{Data rate comparisons between the proposed cocktail BPSK (C-BPSK) and conventional transmission schemes over AWGN channels.}
	\label{fig1}
\end{figure}

In Fig.~\ref{fig1}, the data rates of cocktail BPSK, given by (\ref{eqR1}) and (\ref{eqR2}) with various ratios of $\alpha/\beta$, are compared with conventional BPSK and QPSK schemes over AWGN channels, where $\alpha = \alpha'$ and $\beta = \beta'$.  As expected, with the increase in SNR, the data rate of one-dimensional cocktail BPSK converges to 2 bits/sec/Hz, and the data rate of two-dimensional cocktail BPSK converges to 4 bits/sec/Hz, which pertains to the degrees of freedom in the cocktail BPSK.

\begin{figure}[!t]
\centering
\includegraphics[width=0.47\textwidth]{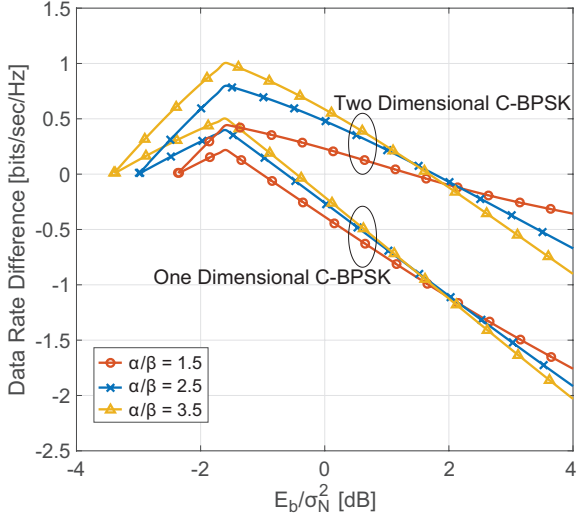}
\caption{Data rate differences between the proposal cocktail BPSK and the AWGN channel capacity achieved by Gaussian-distributed inputs.}
\label{fig2}
\end{figure}

As is shown in this figure, one-dimensional cocktail BPSK achieves higher data rate than conventional BPSK.  Since $\mathbb{R}_2 = 2 \mathbb{R}_1$ herein and the data rate of QPSK is twice that of BPSK, the two-dimensional cocktail BPSK achieves higher data rate than conventional QPSK.  In particular, the competitive advantage of the cocktail BPSK in the low-SNR region comes from the cross power utilization in the demodulation, as justified in (\ref{eqDiff}).

Moreover, with the increase in the ratio of $\alpha/\beta$, the data rate of the cocktail BPSK increases at low SNRs but decreases at high SNRs, which corroborates the analysis in Section \ref{DataRate}.  As manifested in (\ref{eqDiff}), the data rate difference between cocktail BPSK and conventional BPSK increases with the increase in the difference between $\alpha$ and $\beta$, at low SNRs.  However, at high SNRs, the concavity of data rates dominates the performance.  As given in (\ref{eqR1}), there are four items of BPSK data rate functions with different SNRs as the arguments.  The items with larger SNRs bend to the horizontal direction faster than those with lower SNRs.  Thus, the former cannot contribute in the high-SNR region as they do in the low-SNR region.  Consequently, as the SNR increases, the cocktail BPSK performance gain gets reduced in comparison with the AWGN channel capacity gain and, eventually, is saturated by the degrees of freedom.

Specifically at very low SNRs, the data rates of cocktail BPSK are higher than the AWGN channel capacity.  In detail, Fig.~\ref{fig2} reveals the data rate differences between the cocktail BPSK and the AWGN channel capacity, i.e., $\mathbb{R}_1 - C$ and $\mathbb{R}_2 - C$.

Finally, it is observed from Figs.~\ref{fig1} and \ref{fig2} that the data rate of one-dimensional cocktail BPSK and that of the two-dimensional case converge at the same value when $E_b/\sigma_N^2$ decreases to its minimum value for a given ratio of $\alpha/\beta$.  This phenomenon is the same as the way in which the achievable data rates of conventional BPSK and QPSK schemes converge at the point of $E_b/\sigma_N^2 = -1.59$dB, i.e., the fundamental limit in the Shannon capacity theorem to reliably transmit one bit of information per unit resource.  The relation between $E_b/\sigma_N^2$ and the SNR in one-dimensional cocktail BPSK can be expressed as $E_b/\sigma_N^2 = (\rho_z+\rho_x)/\mathbb{R}_1$, and the relation in two-dimensional cocktail BPSK is $E_b/\sigma_N^2 = 2(\rho_z+\rho_x)/\mathbb{R}_2$, where $\rho_z$ and $\rho_x$ are given by (\ref{rhoz}) and (\ref{rhox}), respectively.  Since $\mathbb{R}_2 = 2 \mathbb{R}_1$, the ratio of $E_b/\sigma_N^2$ converges to the same value in both one-dimensional and two-dimensional cases when the SNR goes to zero.

\section{Conclusion}
In this paper, the cocktail BPSK was proposed to layer two independent symbol streams at the transmitter on a non-orthogonal basis and demodulate the layering without inter-stream interference at the receiver, by taking advantage of cross power utilization.  To demonstrate the performance of the proposed scheme, its achievable data rate was established and numerical results were provided for the comparisons with conventional transmission schemes, which not only substantiated the validity of cocktail BPSK, but also offered useful references for the power allocation in the proposed scheme.  Furthermore, it should be noted that the achievable data rate of the cocktail BPSK outperforms the AWGN channel capacity, at very low SNRs.

\appendix

\begin{figure}[!b]
	\centering
	\includegraphics[width=0.3\textwidth]{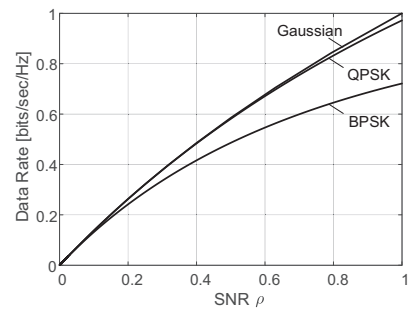}
	\caption{Data rates achieved by the Gaussian-distributed input, QPSK-modulated signals and BPSK-modulated signals over AWGN channels versus the SNR $\rho$.}
	\label{fig3}
\end{figure}

In Fig. \ref{fig3}, the data rates achieved by the Gaussian-distributed input, QPSK- and BPSK-modulated signals are plotted as functions of the decimal SNR $\rho$ over AWGN channels, where all the curves converge to zero data rate as $\rho$ goes to zero.  The following insights are obtained from this figure: I) The difference between achievable data rates of finite-alphabet inputs and the AWGN channel capacity achieved by the Gaussian-distributed input is an infinitesimal. II) The derivatives of all curves involved in this figure are approximately of the same value, i.e., $\log_2 e$.

\end{document}